\documentclass[prl,twocolumn,showpacs]{revtex4}
\usepackage{graphicx}

\newlength{\figwidth}
\begin{document}

%
% frontmatter
%
\title{Numerical Studies of the Compressible Ising Spin Glass}
\author{Adam H. Marshall}
\affiliation{The James Franck Institute and Department of Physics,
   University of Chicago, Chicago, Illinois 60637}
\author{Bulbul Chakraborty}
\affiliation{The Martin Fisher School of Physics, Brandeis University,
   Waltham, Massachusetts 02454}
\author{Sidney R. Nagel}
\affiliation{The James Franck Institute and Department of Physics,
   University of Chicago, Chicago, Illinois 60637}
\date{\today}

\begin{abstract}
We study a two-dimensional compressible Ising spin glass at constant
volume. The spin interactions are coupled to the distance between
neighboring particles in the Edwards-Anderson model with $\pm J$
interactions.  We find that the energy of a given spin configuration is
shifted from its incompressible value, $E_0$, by an amount quadratic in
$E_0$ and proportional to the coupling strength.  We then construct a
simple model expressed only in terms of spin variables that predicts the
existence of a critical value of the coupling above which the spin-glass
transition disappears.
\end{abstract}

% PACS numbers:
% 75.10.Nr Spin-glass and other random models
% 75.40.Mg Numerical simulation studies
% 05.50.+q Lattice theory and statistics (Ising, Potts, etc.)
\maketitle

%
% body
%
Lattice compressibility affects the nature of phase transitions in a
variety of spin systems.  The 2-dimensional (2-D) triangular Ising
antiferromagnet, for example, is fully frustrated and shows no transition
to an ordered state; however, when compressibility is added, this system
exhibits a first-order transition to a striped phase~\cite{chen86,gu__96}.
In the (unfrustrated) Ising ferromagnet, the introduction of
compressibility changes the transition from second- to first-order so that
the onset of nonzero net magnetization is simultaneous with a
discontinuous change in the volume~\cite{bean62,berg76}.  Frustration is
central to the nature of the spin-glass transition as it leads to large
ground-state degeneracies~\cite{bind86,fisc93,pala01,luki04}.  Because
different states with the same spin-glass energy can couple differently to
local lattice deformations, compressibility lifts the ground-state
degeneracy and may dramatically alter the nature of the transition and the
low-temperature spin-glass phase.  Since magneto-elastic effects are
always present to some degree in physical systems, their inclusion in
spin-glass models may help explain some of the outstanding puzzles in
spin-glass experiments~\cite{bitk96}.  We report here the effect of
compressibility on a 2-D spin glass in which the lattice is allowed to
distort locally while the entire system is held at constant volume. 

The ground states of the previously studied compressible systems were
always states that had already been ground states of those same systems
without lattice distortion.  In contrast, we find that as the coupling
between magnetic interactions and lattice distortions is increased, the
constant-volume compressible spin glass prefers spin configurations which
had previously been excited states.  We further find that the critical
region just above the spin-glass temperature is suppressed as the coupling
to lattice distortions increases so that above a certain value of the
coupling, the spin-glass transition is eliminated entirely.  

We study compressible two-dimensional Ising spin glasses on square
lattices with periodic boundary conditions at constant volume.  The
Hamiltonian is
\begin{equation}
   \mathcal{H} = -\sum_{\langle i,j \rangle} J_{ij} S_i S_j
      + \alpha \sum_{\langle i,j \rangle} J_{ij} S_i S_j (r_{ij} - r_0)
      + U_\mathrm{lattice} .
\label{eqn_hamiltonian}
\end{equation}
The first term is the energy of the standard (incompressible)
Edwards-Anderson spin glass, with a sum over nearest neighbors.  The spins
$S_i$ take the values $\pm 1$, and $J_{ij} \in \{\pm J\}$.  The second
term couples the spin interactions to local lattice distortions; $r_{ij}$
is the distance between particles $i$ and $j$, and $r_0$ is the particle
separation of the undistorted lattice.  We consider the coupling to linear
order with coupling constant $\alpha$.  This term allows particles to move
based on their magnetic interactions with their neighbors: a satisfied
interaction (i.e. $J_{ij} S_i S_j = +1$) is enhanced when the particles
move closer together, and an unsatisfied interaction is mitigated when the
particles move apart.  Because the distortions cannot be independent, the
energy of a given configuration will depend on the local arrangement of
satisfied (short) and unsatisfied (long) bonds, thus breaking the
degeneracy inherent to the incompressible model.

The final term in the Hamiltonian represents the energy of the lattice
itself: Hooke's-law springs of uniform spring constant $k$ connect
neighboring particles; springs are also required between next-nearest
neighbors (across the diagonals of the squares) to prevent shear.  All
springs have their unstretched length equal to the natural spacing of
particles on the undistorted lattice.

We identify two important parameters: $\overline\delta = J \alpha / k$ is
a length scale which determines the typical size of the distortions from
the uncompressed locations; the dimensionless quantity $\mu = J \alpha^2 /
k$ gives the average energy of the distortions relative to the original
spin-glass energy.  In our simulations, the parameters are chosen such
that $\overline\delta$ is held fixed at $r_0/10$ while $\mu$ varies over
the range of interest.  The magnetic interaction strength $J$, which we
set to unity, sets the overall energy scale.

We prepare a series of spin states and relax the lattice to its minimum
potential energy via conjugant-gradient minimization~\cite{pres97}.  These
states are generated using standard Monte Carlo techniques on the
incompressible spin-glass Hamiltonian.  For $L = 3,4,5$, we enumerate all
$2^{L^2}$ spin states for each bond realization.  For larger systems, we
acquire data from the ground state up to the highest-entropy states (where
the spin energy is approximately zero) by running at multiple
temperatures.  We average over 100 different bond realizations.

\begin{figure}
\includegraphics[width=\figwidth]{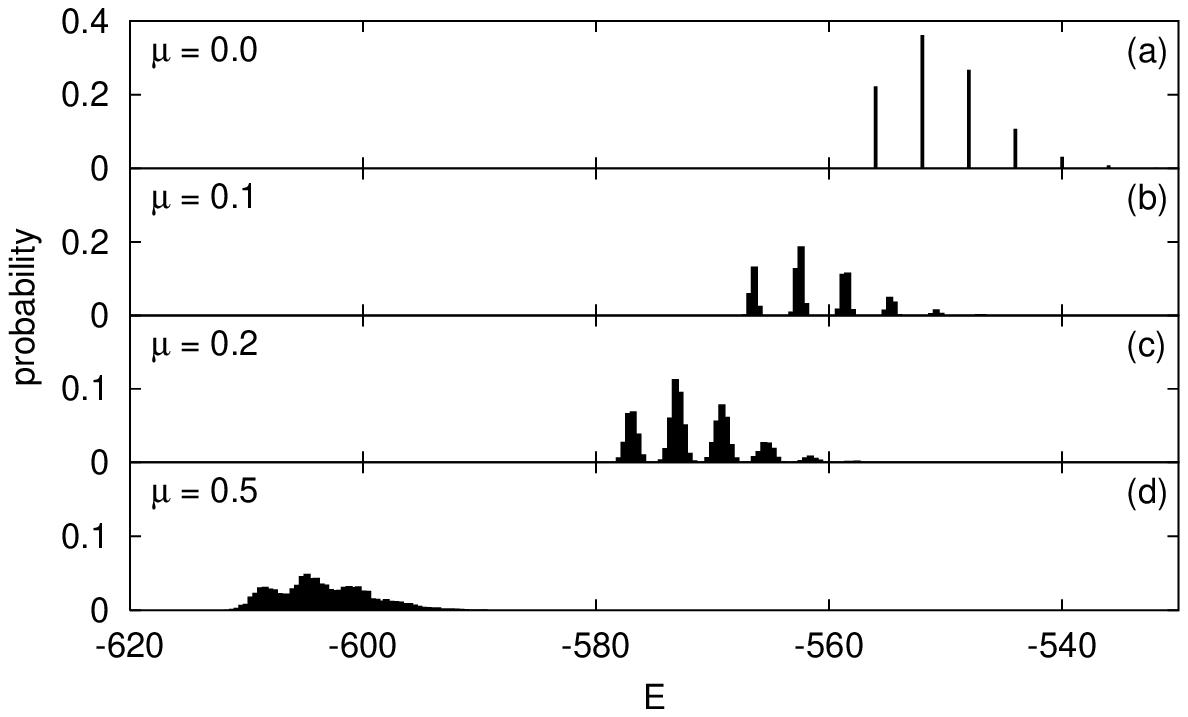}
\includegraphics[width=\figwidth]{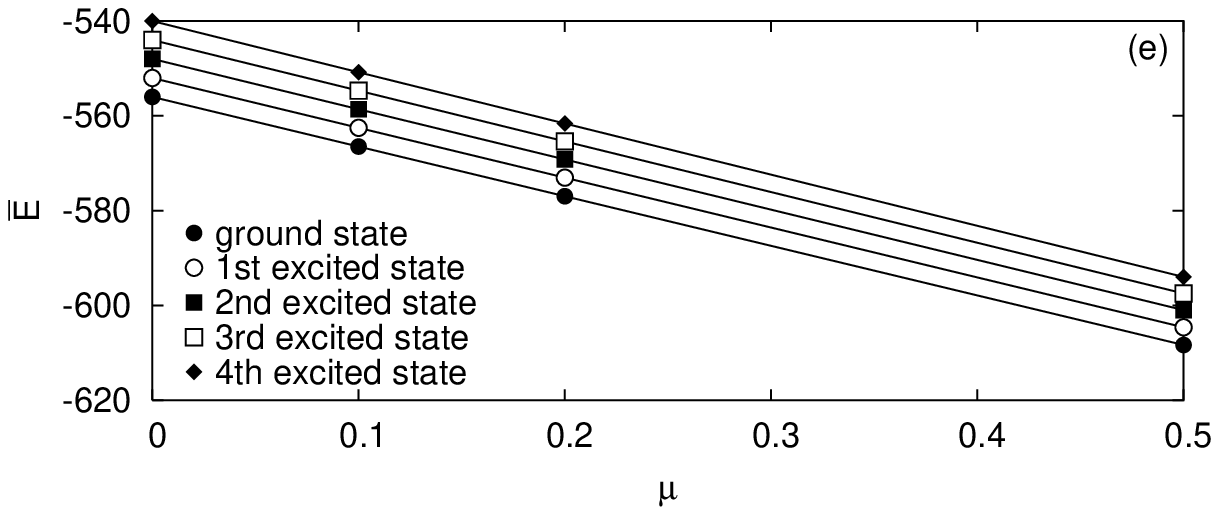}
\caption{(a) The energy spectrum of the incompressible spin glass consists
of a series of delta functions separated in energy by $4J$.  (b)--(d)~As
$\mu$ increases from zero, the spectrum shifts downward in energy, and
each level spreads into a Gaussian-shaped band.  The number of states
under each curve remains constant.  These data are from a single $L=20$
realization at $T=0.65$.  (e)~The average of each band is linear in $\mu$;
the lines associated with different bands have slightly different slopes.}
\label{fig_energy_spread}
\end{figure}

As a function of the coupling, $\mu$, and tmperature, $T$, we examine the
probability, $P(E)$, of finding a state with energy $E$.  Typical results
are shown in Fig.~\ref{fig_energy_spread}.  For the uncoupled spin glass
with $\mu=0$, $P(E)$ is a series of delta functions separated by
$4J$~\cite{luki04} with heights proportional to $g(E) \, e^{-E/T}$, where
$g(E)$ is the density of states (Fig.~\ref{fig_energy_spread}(a)).  As
$\mu$ increases, each initially degenerate level broadens and shifts to
lower energy (Fig.~\ref{fig_energy_spread}(b)--(d)) by an amount
proportional to $\mu$ (Fig.~\ref{fig_energy_spread}(e)).

\begin{figure}
\includegraphics[width=\figwidth]{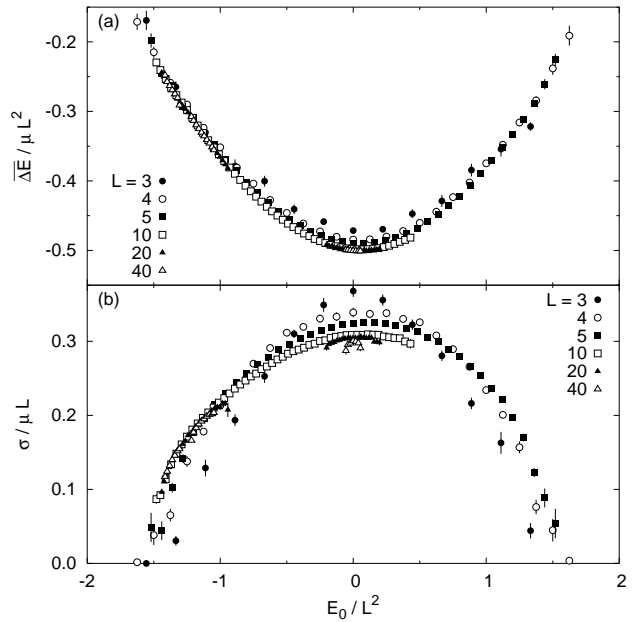}
\caption{(a)~The average energy shift, $\overline{\Delta E} = \overline{E}
- E_0$, divided by $\mu$, plotted as a function of the initial energy;
both axes have been scaled by the system size, $L^2$.  Data for different
system sizes approach a common parabolic curve as $L$ increases.  (b)~The
scaled data for the width of each energy band also approach a common curve
for large system sizes, but the scaling form indicates that the width is
only linear in $L$.  In the thermodynamic limit, the spread of each band
is negligible compared to the average shift in energy.}
\label{fig_size_scaling}
\end{figure}

As $\mu$ is increased, we calculate the energy shift of each
configuration, $\Delta E \equiv E - E_0$, where $E_0$ is the energy of
that configuration at $\mu = 0$.  For each of the original levels (i.e.,
the $\delta$-functions of Fig.~\ref{fig_energy_spread}(a)), we compute its
average shift, $\overline{\Delta E}$, and width, $\sigma$.  Both
$\overline{\Delta E}$ and $\sigma$  are quadratic in $E_0$.  The data for
$\overline{\Delta E}$ can be scaled onto a common curve by dividing by
$L^2$, the number of particles (Fig.~\ref{fig_size_scaling}(a)):
\begin{equation}
   \frac{\overline{\Delta E}}{\mu L^2} \sim  -A + B \left(\frac{E_0}{L^2}
      \right)^2 ,
\label{eqn_energy_size_scaling}
\end{equation}
where $A=0.50$ and $B=0.13$.  The smaller shift for low-energy states than
for those near $E_0=0$ is to be expected since low-energy states, with
many more satisfied than unsatisfied bonds, cannot distort as effectively
as those with roughly equal numbers of each.

The data for the width of each band, $\sigma$, can be scaled onto a common
curve by dividing by $L$ (Fig.~\ref{fig_size_scaling}(b)):
\begin{equation}
   \frac{\sigma}{\mu L} \sim  C - D \left(\frac{E_0}{L^2} \right)^2 ,
\label{eqn_energy_width_scaling}
\end{equation}
where $C=0.31$ and $D=0.090$.  While the energy shift is proportional to
the area, the broadening is proportional only to the linear system size.
In the infinite-size limit, the spreading of each level will be negligible
relative to its energy shift.  Eq.~\ref{eqn_energy_width_scaling} also
implies that the energy gaps between neighboring levels vanish when $\mu
\approx 4 / L$, i.e., for infinitesimal $\mu$ in the infinite-size limit.
The previously discrete spectrum is thus rendered continuous. 

The empirical form of Eq.~\ref{eqn_energy_size_scaling} and the
observation that the width of a level can be neglected in the
thermodynamic limit suggest a simplified model.  Because the shift of
each level, $\overline{\Delta E}$, is simply proportional to ${E_0}^2$, we
replace the last two terms in the original Hamiltonian of
Eq.~\ref{eqn_hamiltonian} by a single term that gives that shift
explicitly:
\begin{equation}
   \mathcal{H}_\mathrm{approx} = -\sum_{\langle i,j \rangle} J_{ij} S_i S_j
      + \frac{\nu}{L^2} \left( \sum_{\langle i,j \rangle}
      J_{ij} S_i S_j \right)^2 .
\label{eqn_hamiltonian_approx}
\end{equation}
Here we have absorbed multiplicative factors into the coupling constant
$\nu \approx \mu / 8$.  The constant offset has been discarded as it does
not affect the thermodynamics.  The first term is again the standard
Edwards-Anderson spin-glass energy.  The second term is the resultant of
the last two terms of Eq.~\ref{eqn_hamiltonian}:  force balance provides a
typical distortion that depends upon the total spin energy, and this
distortion then substitutes into the coupling term, supplying another
factor of the spin energy.  This Hamiltonian contains only spin degrees of
freedom, with four-spin, infinite-ranged interactions.  The long-range
nature of this model reflects the network of interconnected springs in the
original compressible model.

\begin{figure}
\includegraphics[width=\figwidth]{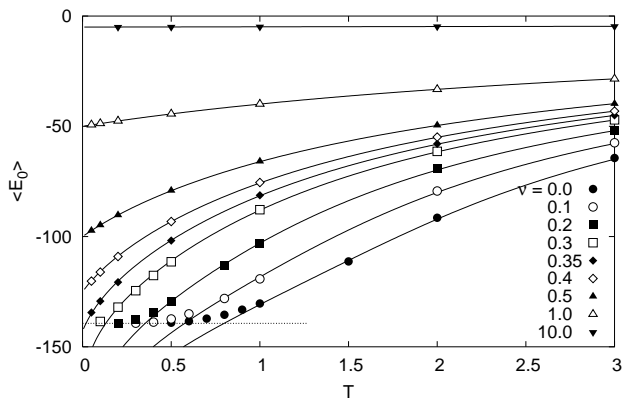}
\caption{Simulation results for $L=10$ systems using the model of
Eq.~\ref{eqn_hamiltonian_approx}; solid lines indicate calculations using
the entropy expression of Eq.~\ref{eqn_model_entropy}.  Below $\nu^*
\approx 0.36$, the average bare spin energy approaches that of the ground
state of the incompressible system (dotted line) as $T$ is lowered.
Above $\nu^*$, however, the zero-temperature value of the average spin
energy is given by Eq.~\ref{eqn_E0_min}.  In this regime, the data may be
predicted quite accurately.}
\label{fig_model_sim}
\end{figure}

The simplified model can be simulated directly and studied analytically.
A straightforward derivation predicts a critical value of $\nu$ at which a
zero-temperature transition takes place.  From the approximate Hamiltonian
of Eq.~\ref{eqn_hamiltonian_approx}, the energy is $E = E_0 +
\frac{\nu}{L^2} {E_0}^2$ (note that because the energy $E_0 \simeq L^2$
the last term does not vanish as $L$ goes to infinity).  The minimum of
this function with respect to $E_0$ gives the level with lowest total
energy and depends on $\nu$ according to
\begin{equation}
   E_{0,\mathrm{min}} = \frac{-L^2}{2 \nu} .
\label{eqn_E0_min}
\end{equation}
For small $\nu$, this minimum lies in the non-physical range below the
ground state $E_{0,\mathrm{ground}}$.  However, when
\begin{equation}
   \nu = \frac{-L^2}{2 E_{0,\mathrm{ground}}} \equiv \nu^* ,
\end{equation}
the first excited state and the ground state of the incompressible system
have the same energy.  For large $L$, the ground state energy per spin,
$E_{0,\mathrm{ground}}/L^2$  has been determined to be approximately
$-1.4$~\cite{wang88}; thus, $\nu^* \approx 0.36$.

Above this value, ground states of the uncoupled system are no longer the
lowest-energy states of the coupled system.  This is seen in simulations
of the simplified model (Fig.~\ref{fig_model_sim}).  As $T$ is lowered to
zero, the thermally averaged value of $E_0$ approaches that of the
original ground state when $\nu < \nu^*$ but approaches the value
predicted by Eq.~\ref{eqn_E0_min} above $\nu^*$.

Because they involve only spin degrees of freedom, the calculations for
this model are standard spin-glass Monte Carlo simulations using the
energy function given in Eq.~\ref{eqn_hamiltonian_approx}.  The presence
of $\nu$ brings neighboring energy levels closer together in energy, and
the large-$\nu$ simulations equilibrate rapidly.  We find that the
$L$-dependence of the simulations is somewhat trivial: when scaled by the
number of particles, computed quantities converge quickly to asymptotic
values as $L$ increases~\cite{mars06}.

The entropy of the $\pm J$ spin glass is well described by a quadratic with
hyperbolic cosine corrections~\cite{mars06}:
\begin{equation}
   S(E_0) = S_0 - S_1 {E_0}^2 - S_2 \cosh(S_3 E_0) .
\label{eqn_model_entropy}
\end{equation}
For $L=10$, typical values are $S_1 = 2.5 \times 10^{-3}$, $S_2 = 8.1
\times 10^{-3}$, and $S_3 = 0.056$.  Using this form, we minimize the free
energy with respect to $E_0$ to obtain the average $E_0$ in the
thermodynamic limit.  The solution can be solved for numerically.
Predicted values for $E_0$ as a function of $T$ are shown as solid lines
in Fig.~\ref{fig_model_sim}; agreement with the data from finite-size
systems is good for large temperatures and values of the coupling above
$\nu^*$.  The breakdown of the prediction near the ground-state energy
results from the lack of a low-energy cutoff in
Eq.~\ref{eqn_model_entropy}.

A transition occurs at $\nu^*$.  For larger values of $\nu$, states of low
total energy are also high-entropy states.  Thus the competition between
energy and entropy, on which the spin-glass transition depends, vanishes.
An order parameter characterizing this new transition is the difference in
spin energy of the level with lowest total energy and of the ground state
of the uncoupled system, $E_{0,\mathrm{min}} - E_{0,\mathrm{ground}}$.
Below $\nu^*$, this quantity is zero; above, it increases as
$\frac{1}{\nu^*} - \frac{1}{\nu}$, i.e.  linearly in $\nu - \nu^*$ near
the transition.

\begin{figure}
\includegraphics[width=\figwidth]{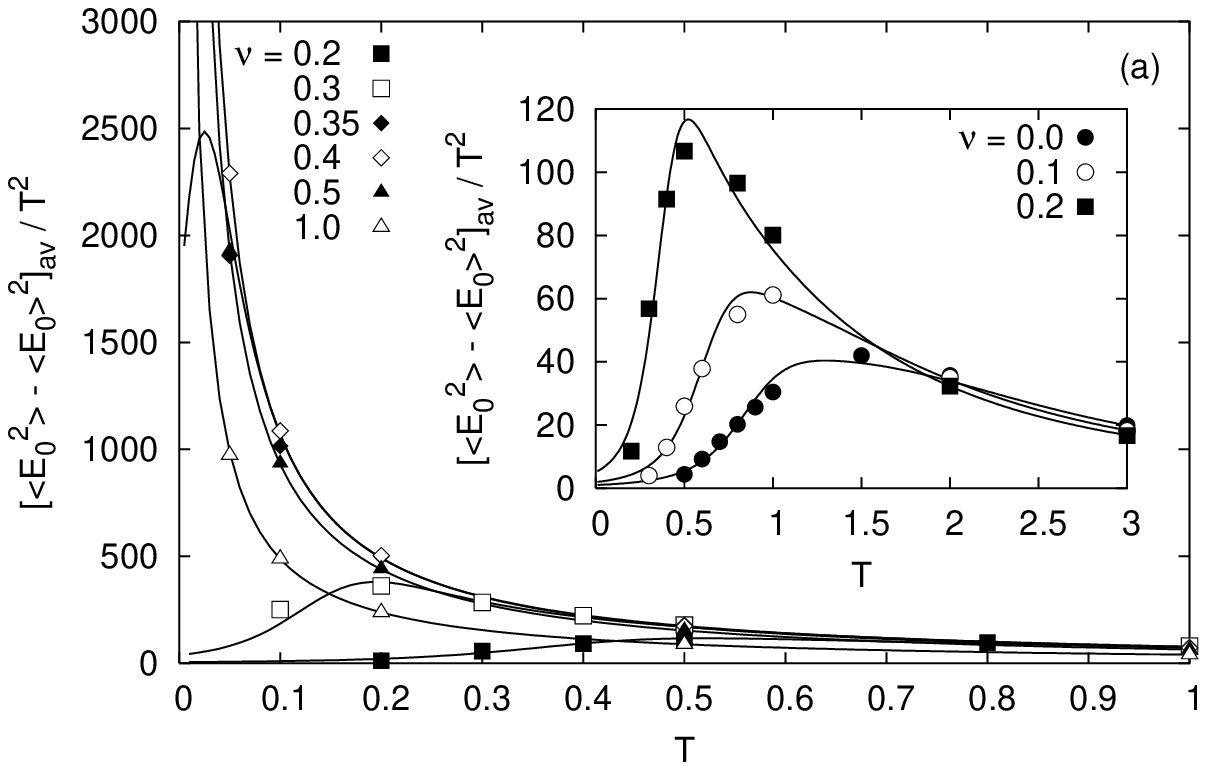}
\includegraphics[width=\figwidth]{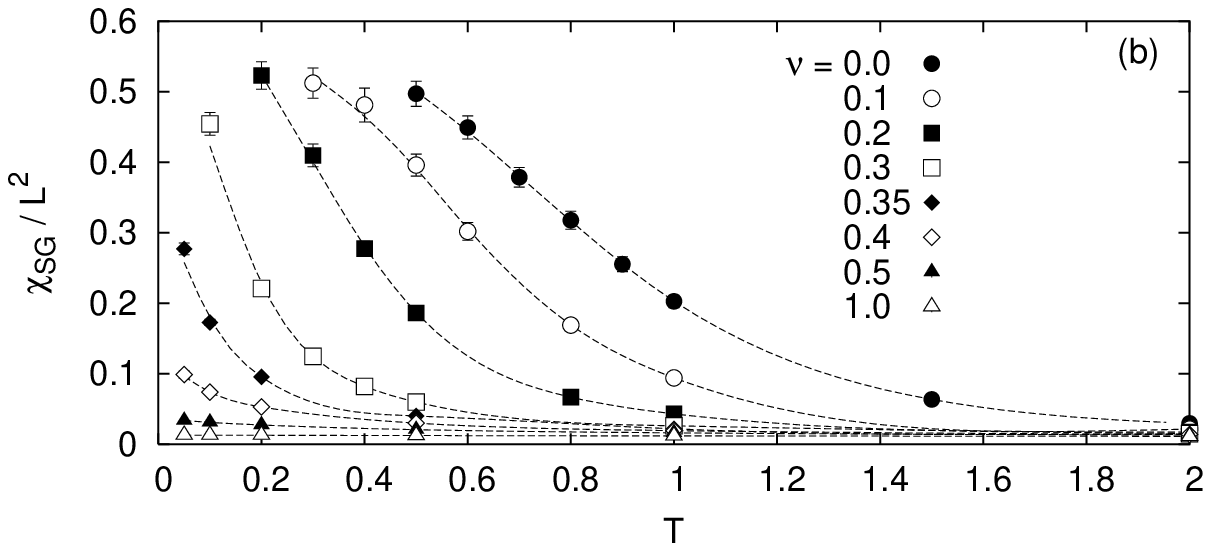}
\caption{(a)~The temperature-reduced fluctuations in $\langle E_0 \rangle$
diverge as $T \rightarrow 0$ when $\nu > \nu^*$.  Solid lines indicate the
curves predicted from free-energy calculations.  (b)~The spin-glass
susceptibility per spin, as a function of temperature.  Dashed lines are
merely guides to the eye.  The sharp rise of $\chi_{SG}$ indicates the
onset of critical behavior; the temperature at which this occurs is pushed
down as $\nu$ increases until there is no longer a critical region.  The
spin-glass phase is thus eliminated.}
\label{fig_model_fluct}
\end{figure}

The temperature-reduced fluctuations of $E_0$ are shown versus $T$ in
Fig.~\ref{fig_model_fluct}(a) where the solid lines are calculations based
on the entropy of Eq.~\ref{eqn_model_entropy}.  As $\nu$ increases, the
maximum value of this quantity increases and shifts to lower $T$, and
there appears to be a zero-temperature divergence that occurs as $\nu
\rightarrow \nu^*$.  This divergence in the fluctuations of the order
parameter occurs due to the energy levels moving closer together, becoming
equal at the critical value $\nu^*$.  The specific heat, which includes
both energy terms in Eq.~\ref{eqn_hamiltonian_approx}, shows no such
behavior and always goes to zero as $T$ is lowered.

The scaling behavior of the spin-glass susceptibility has been used to
determine the temperature and critical exponents associated with the
spin-glass transition~\cite{ogie85b,bhat88,wang88}.
Fig.~\ref{fig_model_fluct}(b) shows the spin-glass susceptibility per spin
versus $T$ for various values of $\nu$.  The increase of this
susceptibility as $T$ is lowered signals the onset of the critical phase
that precedes the spin-glass transition~\cite{ogie85b}.  As $\nu$ is
increased, this critical phase shrinks, implying that the spin-glass
transition is being destroyed. 

In the $T$-$\nu$ plane, there exists an approximate boundary between the
normal paramagnetic phase and the critical phase which signals the approach
of spin-glass behavior.  This boundary can be characterized, as a
function of $\nu$, by the temperature at which the temperature-scaled
order-parameter fluctuations are maximized (see
Fig.~\ref{fig_model_fluct}).  It may also be approximated by the
temperature at which our predicted value of the average spin energy
crosses the ground-state spin energy (see Fig.~\ref{fig_model_sim}).  In
either case the phase boundary terminates at the same critical value
$\nu^*$.

We have limited our study to 2-dimensional systems.  Preliminary results
in 3-D indicate that $\overline{\Delta E}$ and $\sigma$  also depend
quadratically on $E_0$ with the same dependencies on the system size as
were found in 2-D. This would suggest a similar result for how the
spin-glass transition can be destroyed due to compressibility effects in
3-D.  It is important to note that these results rely on the system being
maintained at constant volume.  We expect constant-pressure systems to
behave differently since the low-energy levels should distort more
effectively than higher-energy ones, in contrast to the constant-volume
system.  Finally, since the transition is not dependent upon the discrete
nature of the energy spectrum, its presence is not limited to the $\pm J$
model but rather should be present in Gaussian models as well.

%
% endmatter
%
\begin{acknowledgments}

We thank S.\ Coppersmith, G.\ Grest, S.\ Jensen, J.\ Landry, N.\
Mueggenburg, and T.\ Witten for helpful discussions.  BC acknowledges the
hospitality of the James Franck Institute.  SRN and AHM were supported by
NSF DMR-0352777 and MRSEC DMR-0213745, and BC was supported by NSF
DMR-0207106.
\end{acknowledgments}

%\bibliography{csg_prl}

\end{document}